\documentclass[]{elsarticle}

\usepackage{lineno,hyperref}
\modulolinenumbers[5]

\begin{document}

\begin{frontmatter}

\title{An atomistic perspective of martensite twinning in Iron}

\author{S. Karewar$^{a,b,}$\footnote{Corresponding author.}, A. Elzas, J. Sietsma, M. J. Santofimia}

\address [1]{Department of Materials Science and Engineering, Delft University of Technology, Mekelweg 2, \mbox{2628 CD Delft, The Netherlands}}
\address [2]{Materials Science and Engineering, Institute I, Friedrich-Alexander-Universit$\ddot{a}$t Erlangen-N$\ddot{u}$rnberg}





\date{\today}

\begin{abstract}


The martensitic transformation is one of the most important phenomena in metals science due to its essential contribution to the strength of steels and most engineering alloys. Yet the basic, atomistic mechanisms leading to martensite nucleation and twin morphology are not yet known. A detailed picture in this regard is required if the strengthening effects of martensite are to be properly understood. This work presents molecular dynamics (MD) simulations of the martensitic transformation using a model fcc/bcc semi-coherent interface with Nishiyama-Wasserman orientation relationship. Significant insight into this important phenomenon is detailed in this work which shows that the atomic displacements that cause nucleation and twin morphology formation of the martensitic phase originate at the fcc/bcc interface. The interface facilitates the initial atomic shear during the transformation which in turn causes the stress-induced homogeneous nucleation and twin morphology formation. The understanding of the atomistic processes leading to the twin morphology formation will allow the control of the twinning process for further enhancement of mechanical properties. 


\bigskip
\end{abstract}

\end{frontmatter}


The martensitic transformation is one of the most studied phase transformation phenomenon in metals science because of its scientific and technological importance. It occurs in many material systems and represents the change in crystal structure from one phase to another by a rapid diffusionless transition in crystal structure involving the coordinated movement or shear of atoms during quenching, thermal cooling, or mechanical loading. In Fe and Fe-C alloys it represents a change in crystal structure from face-centered-cubic (fcc) austenite to body-centered-cubic (bcc) or body-centered-tetragonal (bct) martensite which takes place at the speed of sound in the solid, forms a twinned morphology, and controls the mechanical properties \cite{Nishiyamabook,Bhadeshia2017135,Kurdjumow1930,BOGERS1964255,OLSON1972107}. Several theories have been postulated over the years \cite{Nishiyamabook,Bhadeshia2017135} to describe the martensitic transformation, but the atomistic physics involving atomic movements that lead to martensite nucleation, growth, and the formation of the twin morphology are still largely unknown. These theories are based on the observation of the final martensite microstructure in experiments and consider the lattice transformations to match the final observed microstructure of martensite, but do not take into account the exact atomic displacements that take place during the transformation. In addition, since the transformation happens so rapidly, it is impossible to track the atomic motions experimentally. The effect of grain boundaries or interfaces on the transformation mechanisms, which can be significant, are often neglected in the theoretical framework. Interfaces play an important role to control the material properties and phase transformations by affecting the initial defect nucleation and growth at the interfaces, and therefore it is important to account their role on the martensitic phase transformations. 


It is well established in the literature that the martensite has a twinned morphology although the exact atomic mechanism leading to it has not been proposed yet in the literature, to the best of the authors knowledge. As Bhadeshia and Wayman state: \textit{The phenomenological theory of martensitic transformations permits all the crystallographic features of the parent and product phases to be mathematically related, but does not yield detailed information about the mechanism of transformation. The latter depends critically on the structure of the transformation interface} \cite{BHADESHIA20141021}. Molecular dynamics (MD) simulations provide a powerful tool to yield detailed information about the atomistic pathways of the nucleation and growth of the martensitic transformation. In our previous works, we explained the atomic displacements during fcc-to-bcc transformation in single crystal pure Fe and Fe-C simulation systems \cite{KAREWAR201871,KarewarCrystals2019}. The present work represents comprehensive MD simulation results, based on the atomic interactions, explaining the detailed mechanisms of nucleation and growth behaviour of martensite. 

\begin{figure}[htbp]
  \centering
  \includegraphics[scale=0.62]{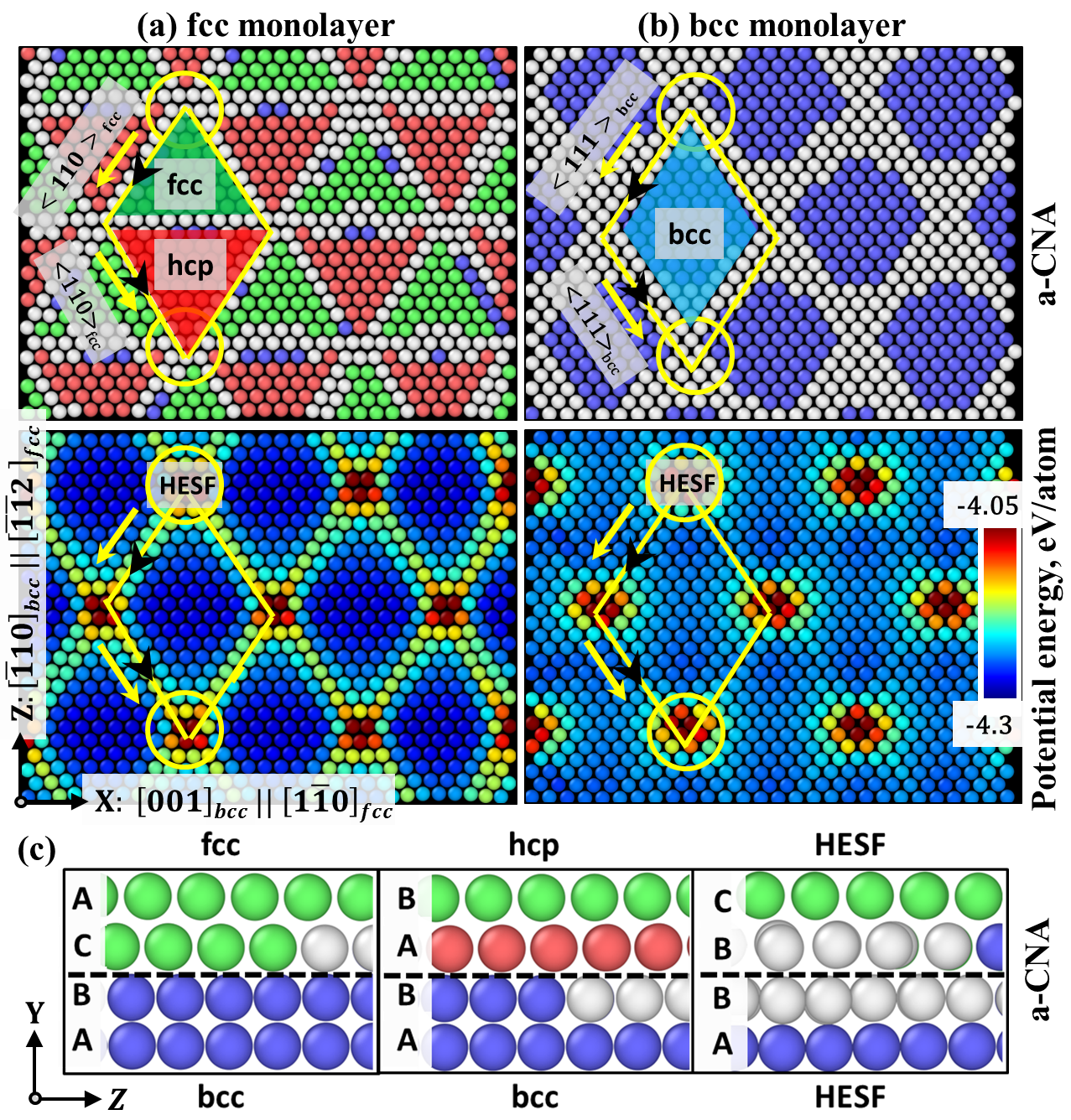}
 \caption{The zoomed in view of two mono-layers that form the unrelaxed NW OR interface in XZ plane. (a) fcc, (b) bcc, coloured by a-CNA (top row) and atomic potential energy (middle row). a-CNA colour coding: blue-bcc, green-fcc, red-hcp, and grey-unidentified. Yellow lines that form rhombohedral indicate the misfit dislocations, black arrows indicate Burgers vectors, and yellow lines with arrows indicate dislocation line directions. The junction between the two misfit dislocations form a node and are labelled as high energy stacking faults (HESF) indicated by circles. The distorted regions (misfit dislocations and HESF regions) are identified as unidentified atoms by a-CNA and have high potential energy. The region inside the rhombohedral geometry is coherent and have low potential energy. On the fcc monolayer side, the coherent region consists of hcp and fcc stacking whereas on bcc monolayer side it consists of entirely bcc regions. (c) The atomic stacking of planes at these fcc, hcp, bcc, and HESF regions at the coherent and distorted regions projected in YZ plane. Two layers each from fcc and bcc phases are shown and dashed line indicates the interface. The label on top and bottom of the rectangle indicates the stacking on the fcc and bcc sides, respectively, after joining in NW OR.}
  \label{fig:simulation_systems}
\end{figure}

A model bicrystal fcc/bcc Fe simulation system with a semi-coherent interface with Nishiyama-Wasserman (NW) orientation relationship is used to understand the atomic physics (See supplementary material for details of simulation system orientation and dimensions). The atomic structure of the NW interface consists of coherent and distorted regions determined by the lattice constants of the two phases (Fig. \ref{fig:simulation_systems}(a-b)). The coherent regions are identified as fcc, bcc, and hcp phases in rhombohedral geometry (Fig. \ref{fig:simulation_systems}(a-b)). The coherent regions have minimal lattice mismatch strains between the two phases when joined in NW orientation. This results in stable regions with low atomic potential energy. The lattice mismatch is accommodated in the distorted regions which therefore have high atomic potential energies (Fig. \ref{fig:simulation_systems}(a-b)). The distorted regions are indicated by yellow lines and circles (Fig. \ref{fig:simulation_systems}(a-b)). The distorted  regions act as discrete defects in the form of misfit dislocations \cite{HOWE2009792} and the intersection of the two misfit dislocations form a node and are termed as high energy stacking faults (HESF). The misfit dislocations have a screw character with Burgers vector $\left\langle 110 \right\rangle$ and are glissile within the interface plane along the $\left\lbrace 111\right\rbrace  \left\langle 110\right\rangle $ fcc slip system ( see supplementary material for dislocation line direction and Burgers vector of these misfit dislocations). Thus the movement of the misfit dislocations with screw character within the interface plane make the interface glissile, the mobility out of the plane or normal to the interface is achieved by the atomic displacements created by the shear of the misfit dislocations which will be discussed later in this work. 

\begin{figure}[htbp]
  \includegraphics[scale=0.48]{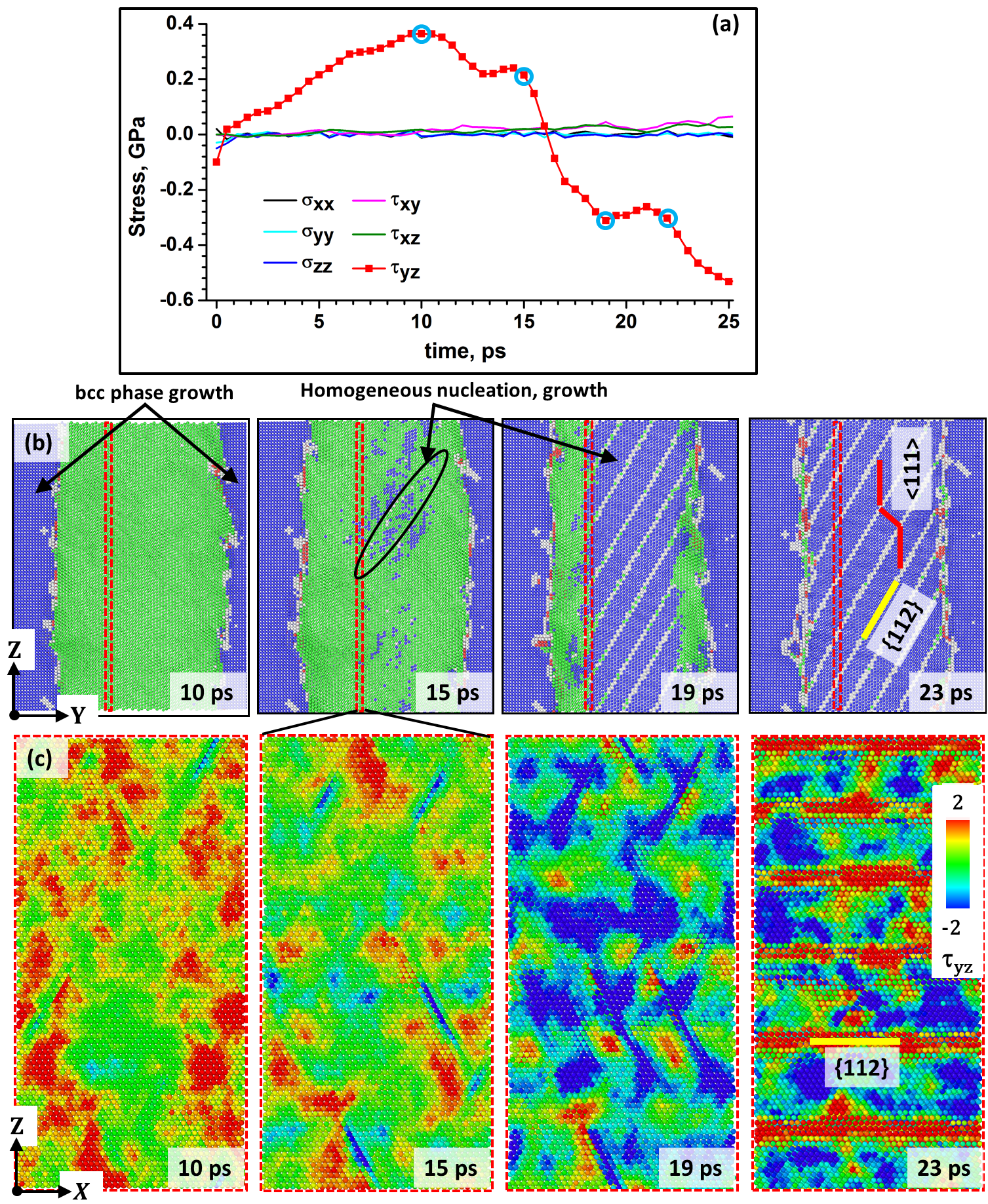}
 \caption{(a) The evolution of the six components of the stress tensor as a function of simulation time. Blue circles indicate the time at which atomic structure and atomic stresses are shown in (b-c). The evolution of atomic structure at 100 K coloured by a-CNA in Y-Z plane (b), and  coloured by shear stress $\tau_{yz}$ (GPa) as per the colour bar (c). The entire fcc phase and only part of the bcc phase near the interface region are shown to focus on the fcc-to-bcc transformation in (b-c). (b) The grey atoms in  between two bcc phase crystallites signify the twin boundaries (TB). The red solid lines drawn on the bcc phase crystallites indicate the $\left\langle 111 \right\rangle_{bcc} $ direction. These lines are mirror images of each other across the TB, which indicates the twinned morphology of the newly formed bcc phase. The yellow line indicates the $\left\lbrace 112 \right\rbrace $ bcc plane which is the habit plane between the original fcc austenite and final formed martensite. The dashed rectangles in (b) indicate the two atomic layers at which $\tau_{yz}$ stresses are shown in XZ plane in (c).}
  \label{fig:Nodefects_Xmation}
\end{figure}


The evolution of six components of the stress tensor in the entire simulation system, and the evolution of the atomic structure and atomic stresses in the fcc phase at 100 K are tracked to understand the fcc-to-bcc transformation mechanism (Fig. \ref{fig:Nodefects_Xmation}(a-c)). At 10 ps, it is observed that only $\tau_{yz}$ stresses have magnitudes that deviate from 0 MPa (Fig. \ref{fig:Nodefects_Xmation}(a)). The magnitudes of the other stress components fluctuate around $\pm$0 MPa. The shape and volume of the simulation box are allowed to change independently along three crystallographic directions which leads to hydrostatic stress components and two of the shear stress components, $\tau_{xy}$ and $\tau_{xz}$, being close to 0 MPa. Only $\tau_{yz}$ shear stresses increase and reach a maximum of 365 MPa as simulation progresses. At this timestep $t = $10 ps, the growth of the bcc phase at two interfaces is observed (Fig. \ref{fig:Nodefects_Xmation}(b)).

These high shear stresses are generated as a direct result of $\Delta{G}_{bcc-fcc} < 0$ at 100 K for this potential \cite{Engin200816}. The free energy difference ($\Delta{G}_{bcc-fcc}<0$) is the thermodynamic driving force for this transformation. This leads to highly unstable fcc phase and high shear stresses are generated within the fcc phase that can cause atomic shear leading to fcc-to-bcc phase transformation. Therefore these high shear stresses, $\tau_{yz}$, are relieved by the atomic shear or atomic displacement within the fcc phase that causes the homogeneous nucleation and growth of bcc phase at 15 ps (Fig. \ref{fig:Nodefects_Xmation}(a-b)). $\tau_{yz}$ reduces to 214 MPa at this time step. Note that the $\tau_{yz}$ shear stresses are not externally applied but thermally induced because of the free energy balance criteria as described. The generation of high stresses during martensitic transformation was observed in previous MD studies of single crystals as well \cite{Wang2014399,Sandoval2009bain}.

At increased simulation times, 15 and 19 ps, the homogeneous nucleation and growth of bcc phase at several places can be seen within the fcc phase. The homogeneous nucleation of one martensite embryo leads to a burst phenomenon called as \textit{autocatalysis} wherein the nucleation of single variant of martensite nucleates many martensite embryos within fcc phase. Since $\tau_{yz}$ is the only component of the shear stress causing the transformation, the atomic shear happens only in YZ plane. At 23 ps, the growth of the homogeneously nucleated bcc phase consumes the entire fcc phase and fcc-to-bcc transformation is complete (Fig. \ref{fig:Nodefects_Xmation}(b)). The original fcc/bcc interface moves by  eight to ten atomic layers within the fcc matrix from both sides and the rest of the fcc phase is consumed by the growth of the bcc phase nucleated homogeneously within the fcc matrix. The thermally induced martensite transformation (TIMT) takes place in the present simulations without the presence of any external stresses \cite{Sakamoto2002}. The martensite embryo forms by the generation of thermally induced shear stresses $\tau_{yz}$ that can deform the surrounding untransformed parent fcc matrix elastically. On the other hand, the parent fcc matrix opposes this transformation by generating elastic back stress. 

The growth of different martensite embryos lead to twin morphology formation. These twins are indicated by red lines which are mirror images of each other in figure \ref{fig:Nodefects_Xmation}(b) at 23 ps. The twin morphology formation also leads to an alternating stress state across the TB (Fig \ref{fig:Nodefects_Xmation}(c)). The nucleation and growth of the bcc phase within the fcc matrix results in a twinned morphology (Fig. \ref{fig:Nodefects_Xmation}(d)), which is also observed experimentally in Fe alloys \cite{Bhadeshia2017135,Sibhata2008twin,ZHANG2016169}. MD simulations offer the advantage over experiments that the exact atomic displacements during the martensitic transformation can be analysed. 

Since this nucleation within the fcc phase is caused by the thermally induced shear stresses, we refer to this as stress-induced homogeneous nucleation. In the literature, this nucleation has been referred to as homogeneous nucleation but the exact atomic displacements leading upto the transformation are not analysed. Nucleation is defined as a localized formation of the distinct thermodynamic phase. There are different kinds of nucleation. Heterogeneous nucleation is the nucleation and growth of the embryo at heterogeneous sites such as grain boundaries, shear bands, intersecting stacking faults, or surfaces \cite{Christianbook}. Homogeneous nucleation, occurs within the bulk parent phase without any pre-existing preferential nucleation sites. Nucleation can also be divided as stress and strain induced nucleation. Stress-induced nucleation occurs on the same sites which trigger the spontaneous transformation on cooling but assisted by the thermodynamic effect of stress, whereas strain induced nucleation implies the production of new nucleation sites by plastic deformation \cite{Olson1975kinetics}. The occurrence of homogeneous nucleation has often been overlooked in the experiments because of the assumption that the barrier for heterogeneous nucleation is less than that for homogeneous nucleation. In addition, it is difficult to observe homogeneous nucleation in-situ in experiments. Here we presented the evidence for the stress assisted homogeneous martensitic nucleation within the fcc matrix.

\begin{figure}[htbp]
  \centering
  \includegraphics[scale=0.6]{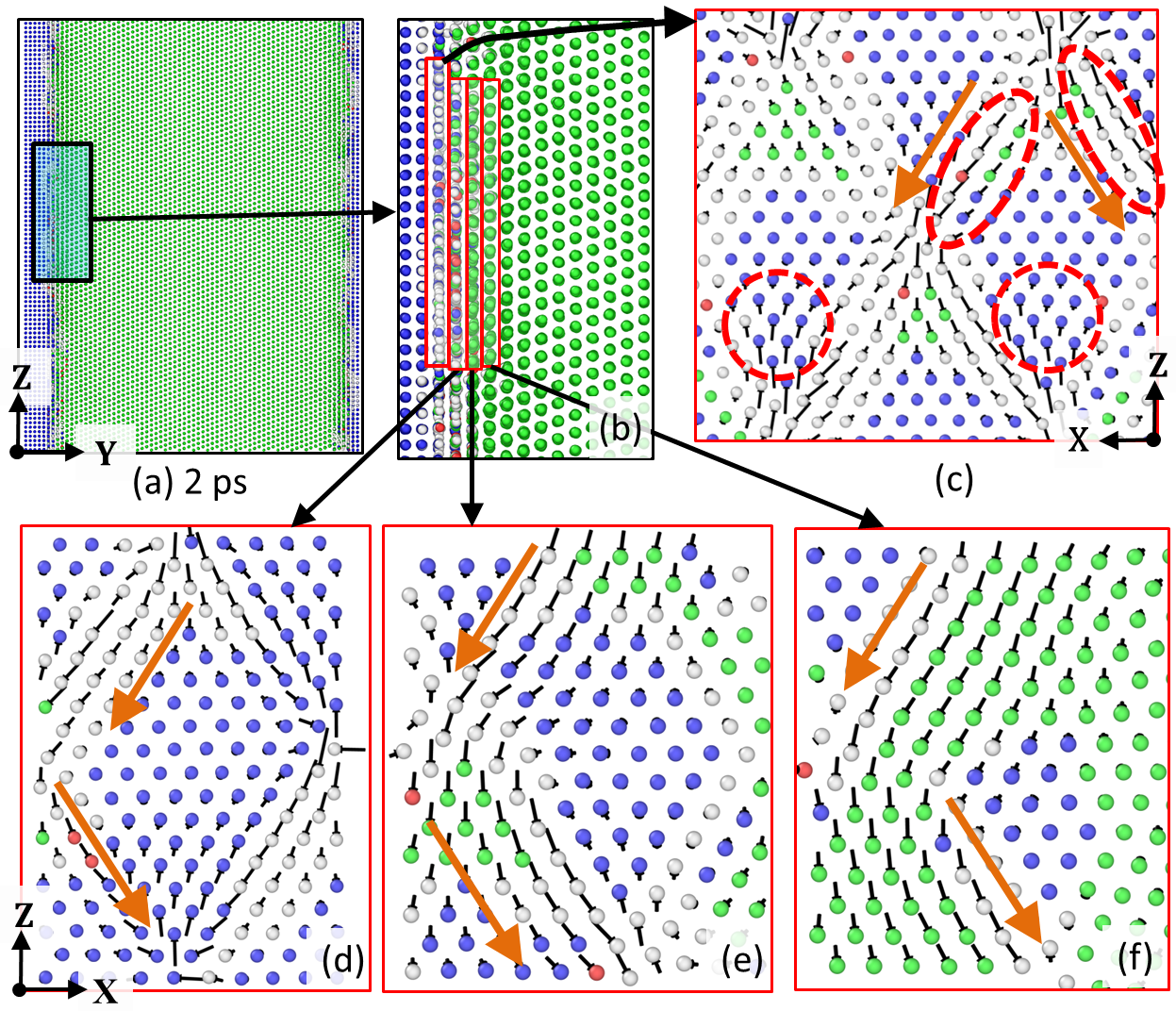}
 \caption{The atomic displacements in four atomic planes at and close to the interface that lead to the twinned morphology at 2 ps. The atomic displacements are calculated with respect to the initial positions of the atoms and are scaled by a factor of three for better visibility. The displacements are shown by black arrows at the head of the atoms and they indicate the original position from where atoms are displaced. Orange arrows indicate the $\left\lbrace 111 \right\rbrace \left\langle 110 \right\rangle $ fcc slip system and are marked as a guide to indicate the shear direction of the group of atoms. (a) The microstructure in the simulation system at 2 ps. (b) Part of the bcc phase close to the interface and the entire fcc phase. Four slices on four successive fcc atomic layers at and next to the interface are taken to analyse the atomic displacements in (c-f) along X-Z plane which is the $\left\lbrace  111 \right\rbrace$ fcc plane. In the fcc plane at the interface (c) the coherent atoms in the fcc phase displace along Z $\Vert \left\langle 112 \right\rangle_{fcc} $ direction to transform to the bcc phase as per BB/OC model (indicated by circles). At the same time the misfit dislocations glide on the $\left\lbrace  111 \right\rbrace_{fcc}$ plane along two $\left\langle 110 \right\rangle_{fcc} $ slip directions in a zig-zag manner (indicated by orange arrows and dashed ellipse). In the next fcc layer (d) the atomic  displacements are prominently on the two $\left\lbrace  111 \right\rbrace\left\langle 110 \right\rangle_{fcc}$ fcc slip systems (indicated by orange arrows). The same atomic displacement pattern is also observed in the next fcc layers (e,f).}
  \label{fig:Twin_origin}
\end{figure}

To understand the role of atomic structure of NW OR interface on the fcc-to-bcc transformation, we look at the atomic displacements that originate at the interface and are at the origin of the twin morphology formation. Although the twin morphology was formed during the nucleation and growth of the bcc phase within the fcc matrix, we observed that the atomic displacements that cause it originate at the interface. Therefore it is important to understand the initial atomic motion at the interface that sets in the fcc-to-bcc transformation. The atomic displacements are analysed at 2 ps on the four fcc layers which are at and next to the interface, and are oriented parallel to the interface (Fig. \ref{fig:Twin_origin}(a-b)). In the fcc monolayer at the  interface, (Fig. \ref{fig:Twin_origin}(c)), two types of atomic shears are observed: first the coherent atoms in the fcc phase displace along the $\left\langle 112 \right\rangle$ direction to transform to bcc phase (indicated in orange circles), according to the Burgers-Bogers-Olson-Cohen (BB/OC) model \cite{OLSON1972107}. This model is well explained in previous MD studies of the fcc-to-bcc transformation mechanism \cite{KAREWAR201871,KarewarCrystals2019,Ou2016}. Secondly misfit dislocations glide on the $\left\lbrace 111 \right\rbrace$ fcc plane along two $\left\langle 110 \right\rangle$ slip directions, in a zig-zag manner (indicated by ellipse in Fig. \ref{fig:Twin_origin}(c)). As discussed previously, these misfit dislocations are high-energy distorted regions and are glissile within the interface plane, and therefore they glide on the $\left\langle 110 \right\rangle$ $\left\lbrace 111 \right\rbrace$ fcc slip system. This leads to the growth of bcc phase at the fcc/bcc interfaces. 

The atomic displacements on the next fcc atomic layer (Fig. \ref{fig:Twin_origin}(d)) are on the $\left\lbrace  111 \right\rbrace$ fcc plane and along two $\left\langle 110 \right\rangle$ fcc directions in a zig-zag manner, and they follow the atomic displacements created by the shearing of the misfit dislocations on the previous fcc layer. The same atomic displacement pattern is observed in the next fcc atomic layers (Fig. \ref{fig:Twin_origin}(e-f)) where atoms follow the atomic displacements on the fcc slip system of the previous two  layers. This atomic displacement is carried forward successively within the fcc matrix on the other fcc atomic planes as growth of the bcc phase progresses. Thus atomic shear normal to the interface is achieved by the atomic displacements created by the shearing of the misfit dislocations. The atomic displacements along two $\left\langle 110 \right\rangle $ directions create alternating twins. This mechanism is similar to the inhomogeneous shear or the lattice invariant deformation from the phenomenological theory of the martensite transformation \cite{BHADESHIA20141021}. The twinned morphology of the martensite is formed to reduce the accumulation of the transformation strain at the interface over long distances. The same type of atomic displacements are observed close to the interface at the right hand side of the simulation system on $\left\lbrace  111 \right\rbrace$ slip planes but along upwards $\left\langle 110 \right\rangle$ directions (Fig. \ref{fig:Twin_origin}(e)).  


\begin{figure}[htbp]
  \centering
  \includegraphics[scale=0.55]{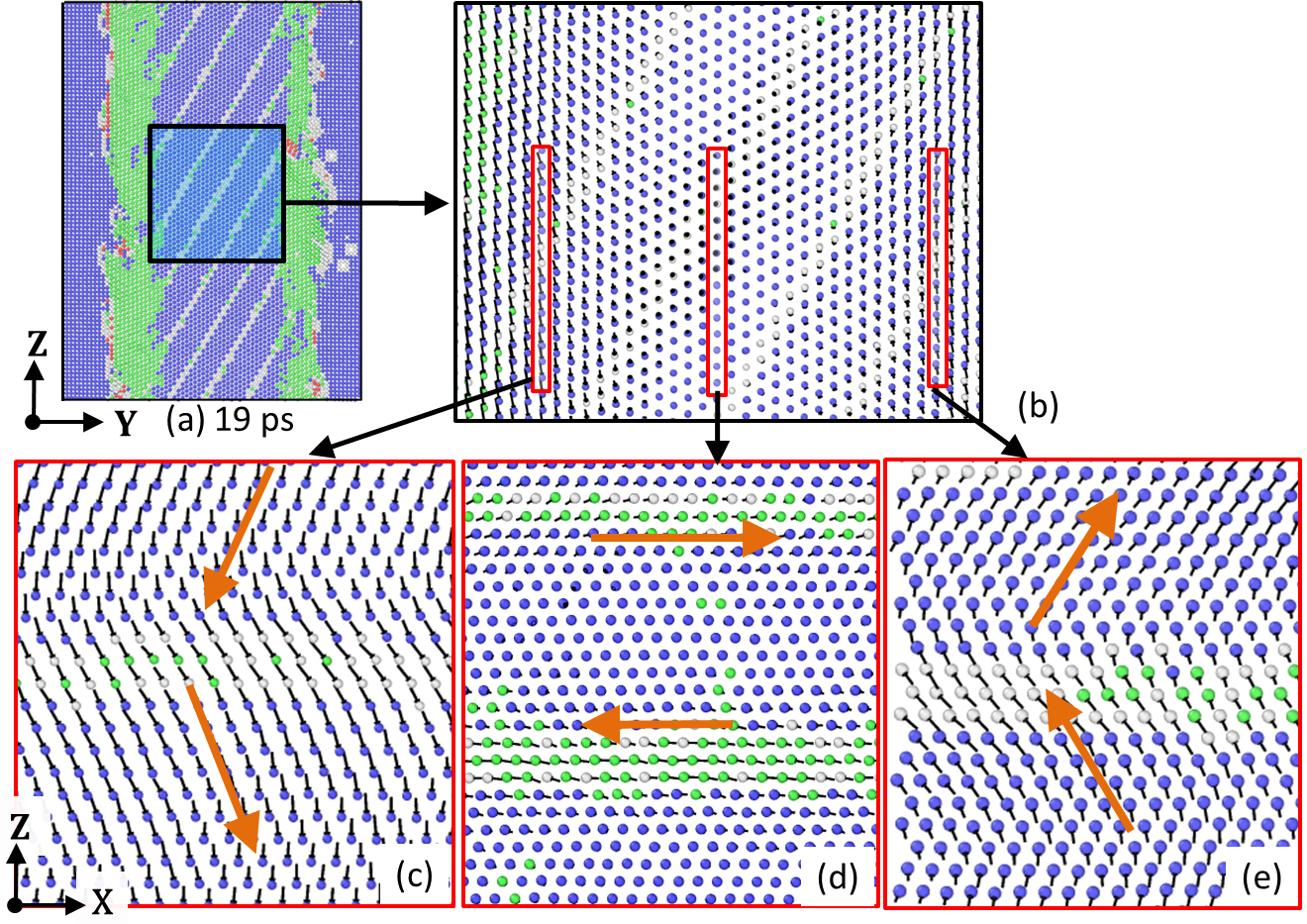}
 \caption{The precise atomic displacements during the nucleation and growth of the bcc phase are traced. (a) The simulation system at 19 ps. (b) A zoomed in view of a monolayer of atoms. The arrows at the atom heads show the original positions from where they are displaced. Orange rectangles indicate three different regions of directions of atomic displacements. These regions are shown in (c-e) along the X-Z projection. In these three regions, the atomic displacements are seen on the same $\lbrace 111\rbrace$ plane but along different $\left\langle 110 \right\rangle$ directions in a zig-zag manner, (c) downwards along Z, (d) right-left along X, and (e) upwards along Z, respectively. The displacement along different $\left\langle 110 \right\rangle$ directions on the same slip plane lead to the twinned morphology. Orange arrows indicate the $\left\lbrace 111 \right\rbrace \left\langle 110 \right\rangle $ fcc slip system and are marked as a guide to indicate the shear direction of the group of atoms.}
  \label{fig:No_defects_disp_growth}
\end{figure}

To understand the formation of twin morphology of bcc phase, the atomic displacements within the fcc matrix are analysed at 19 ps  with respect to the original atomic positions (Fig. \ref{fig:No_defects_disp_growth}). At 19 ps, the homogeneously nucleated bcc phase grows within the fcc matrix to form a twinned morphology (Fig. \ref{fig:No_defects_disp_growth}(a)). The atomic displacements in the three distinct regions are now analysed (Fig. \ref{fig:No_defects_disp_growth}(b)). In the region at the left (Fig. \ref{fig:No_defects_disp_growth}(c)), the atoms have displaced downwards along two different $\left\langle 110 \right\rangle_{fcc} $  directions in a zig-zag manner. The  $\left\langle 110 \right\rangle_{fcc} $ directions correspond to $ \left\langle 111 \right\rangle_{bcc} $ directions. In the region at the right (Fig. \ref{fig:No_defects_disp_growth}(e)), the atoms have displaced upwards along Z direction. The displacements of some of the atoms in the center of fcc phase (Fig. \ref{fig:No_defects_disp_growth}(d)) are towards right and left along two $\left\langle 110 \right\rangle $ fcc directions, whereas for rest of the atoms no displacements are observed. 


This shearing of the atomic planes on the same $\left\lbrace  111 \right\rbrace$ fcc slip plane but along different $\left\langle  110 \right\rangle$ directions in a zig-zag manner creates the twinned morphology in the final bcc phase. The atoms which shear in a coordinated manner along the same $\left\langle  110 \right\rangle$ direction form one bcc twin lamella, and the intersection region between adjacent twin lamellae form the twin boundaries. The zig-zag atomic displacements which result in twin morphology of bcc martensite originates from the glissile misfit dislocations with screw character at the interface. Thus here we observe that the interface structure affects the formation of the final morphology after the transformation. 


The coordinated displacements (Fig. \ref{fig:No_defects_disp_growth}(c) and (e)) are along two opposite $\left\lbrace 111 \right\rbrace \left\langle 110 \right\rangle $ fcc or $\left\lbrace 110 \right\rbrace \left\langle 111 \right\rangle $ bcc slip systems, downwards and upwards respectively, which transforms the fcc structure into bcc. In the center of fcc phase (Fig. \ref{fig:No_defects_disp_growth}(d)), the atoms that form the twin boundary in the transformed martensite displace towards right and left along $\left\langle 110 \right\rangle_{fcc}$ direction. Based on the atomic displacements in fcc phase and their correspondence to bcc slip system, the orientation relationship (OR) during the fcc-to-bcc transformation is written as,\\
$\left( 111 \right)_{fcc} || \left( 110 \right)_{bcc}$, $\left[ 10\bar{1} \right]_{fcc} || \left[ 11\bar{1} \right]_{bcc} $, \\
$\left( 111 \right)_{fcc} || \left( 110 \right)_{bcc}$, $\left[ \bar{1}0\bar{1} \right]_{fcc} || \left[ \bar{1}1\bar{1} \right]_{bcc} $, \\
$\left( 111 \right)_{fcc} || \left( 110 \right)_{bcc}$, $\left[ 101 \right]_{fcc} || \left[ 111 \right]_{bcc} $, \\
$\left( 111 \right)_{fcc} || \left( 110 \right)_{bcc}$, $\left[ \bar{1}01 \right]_{fcc} || \left[ \bar{1}11 \right]_{bcc} $, \\

This signifies the Kurdjumov-Sachs (KS) type OR for the fcc-to-bcc martensitic transformation. All these displacements take place in a coordinated manner. The magnitude of the displacements varies from 0 \AA \hspace{0.05cm} to 1.75 \AA \hspace{0.05cm}. The coordinated displacement of group of atoms with a magnitude less than one interatomic distance ($<$ 2.87 \AA \hspace{0.05cm} in bcc or $<$ 3.61 \AA \hspace{0.05cm} in fcc) suggests the martensitic nature of the transformation. Thus, coordinated atomic displacements of martensitic nature are at the origin of the fcc-to-bcc transformation. Since $\tau_{yz}$ is the only shear component driving the atomic shears during transformation in the present simulation geometry and prescribed boundary conditions, only the martensite variants are formed which lie on the $(110)_{bcc}$ plane but the slip directions within these bcc planes are different. 


In conclusion, simulations on an fcc/bcc bicrystal with NW interface orientation relationship show that the growth of the bcc phase at the interfaces leads to stress-induced homogeneous nucleation within the fcc phase. It is observed that the interface structure aids the nucleation of the martensite phase within the fcc matrix, and also controls the atomic displacements and growth morphology within the fcc matrix. The glissile misfit dislocations at the fcc/bcc interface initiate the atomic displacements in a zig-zag manner that facilitates the formation of the twinned morphology. The stress-induced martensite nucleation is observed within the fcc matrix and the growth happens by the KS mechanism. The final  martensite is twinned which is consistent with the experimental observations. 

The atomistic mechanisms discussed in the present work are based on the study of a small simulation system with periodic boundary conditions. The constraint on the system forces the system to transform with the mechanisms as described. In a less constrained system, such as a bigger simulation system (as described in the supplementary material), atoms can also move normal to the interface plane because of the presence of the other shear components ($\tau_{xy}$ and $\tau_{xz}$). This leads to the formation of smaller secondary twins within the bigger sized primary twins. This autocatalytic process or burst phenomena was also observed experimentally in Fe-Ni steels \cite{Yu1989} where a large number of martensite plates were formed simultaneously. 

The analysis presented in this work can nevertheless be used to understand the mechanisms in the larger systems, as demonstrated for the bigger simulation system sizes. The mechanisms identified in this work are for a pure Fe model. Steel, used in the real conditions, contains alloying elements in Fe. Although these alloying elements can have a strong effect on the morphology of martensite, the fact that the martensite morphology formed in the present work is in good agreement with the experimental observations suggests that the mechanisms described in this work based on pure Fe interatomic interaction can yield detailed information of the basic atomistic phenomena of nucleation and twin formation observed in steels.

The exact transformation mechanisms of stress-induced nucleation and twin morphology formation triggered by the interface structure are shown for the first time. We conclude that the analysis of the atomistic physics, such as discussed in this work can improve our understanding of the dependence of the   fundamental mechanisms on the interface structure. This can be useful in future to tailor materials with optimal properties and control the material transformation behaviour. 


\section*{Acknowledgements}
This work received the funding from the European Research Council under the European Union’s Seventh Framework Programme FP7/2007–2013/ERC grant agreement number [306292]. We acknowledge Prof. Barend Thijsse for useful discussions on this work. 

\section*{Contributions}
Authors declare no conflict of interest. SK, JS, and MS designed the research. SK and AE prepared the initial simulation systems. SK carried out the MD simulations and wrote the initial manuscript. All authors contributed to the analysis of the simulations and preparation of the final manuscript.

\bibliographystyle{ActaMatnew-2}

\bibliography{bibliography}

\end{document}